# Polarization Characteristics of Zebra Patterns in Type IV Solar Radio Bursts

K. KANEDA[1*], H. MISAWA[1], K. IWAI[2], F. TSUCHIYA[1], T. OBARA[1], Y. KATOH[3], and S. MASUDA[4]

[1] Planetary Plasma and Atmospheric Research Center, Tohoku University, Sendai, Miyagi 980-8578, Japan;

[*] k.kaneda@pparc.gp.tohoku.ac.jp

[2] National Institute of Information and Communications Technology, 4-2-1, Nukui-Kitamachi, Koganei, Tokyo 184-8795, Japan

[3] Department of Geophysics, Graduate School of Science, Tohoku University, Sendai, Miyagi 980-8578, Japan

[4] Institute for Space – Earth Environmental Research, Nagoya University, Nagoya, Aichi 464-8601, Japan

## ABSTRACT

The polarization characteristics of zebra patterns (ZPs) in type IV solar bursts were studied. We analyzed 21 ZP events observed by the Assembly of Metric-band Aperture Telescope and Real-time Analysis System between 2010 and 2015 and identified the following characteristics: a degree of circular polarization (DCP) in the range of 0%–70%, a temporal delay of 0–70 ms between the two circularly polarized components (i.e., the right- and left-handed components), and dominant ordinary-mode emission in about 81% of the events. For most events, the relation between the dominant and delayed components could be interpreted in the framework of fundamental plasma emission and depolarization during propagation, though the values of DCP and delay were distributed across wide ranges. Furthermore, it was found that the DCP and delay were positively correlated (rank correlation coefficient $R = 0.62$). As a possible interpretation of this relationship, we considered a model based on depolarization due to reflections at sharp density boundaries assuming fundamental plasma emission. The model calculations of depolarization including multiple reflections and group delay during propagation in the inhomogeneous corona showed that the DCP and delay decreased as the number of reflections increased, which is consistent with the observational results. The dispersive polarization characteristics could be explained by the different numbers of reflections causing depolarization.

## 1. INTRODUCTION

Solar radio bursts are produced by non-thermal electrons that are accelerated in the corona through incoherent mechanisms, such as gyrosynchrotron emission, and coherent mechanisms, such as plasma emission. In the metric wavelength range, the majority of such bursts are produced through the plasma emission mechanism. Since the emission is generated at the local plasma frequencies in the source region, its form in the dynamic spectrum reflects the motion of the emission source. For example, the rapid frequency drift of type III bursts corresponds to the motion of electron beams propagating along open magnetic field lines; in contrast, the slow frequency drift of type II bursts corresponds to magnetohydrodynamic shocks associated with coronal mass ejections (Mclean & Labrum 1985).

In the plasma emission mechanism, radio bursts are generated through several radiative processes, such as acceleration of electrons, excitation of electrostatic plasma waves, conversion of electrostatic waves into electromagnetic waves, and propagation through the corona and interplanetary space. The observed characteristics of bursts can be influenced by modulations and inhomogeneities in these processes. As a result, a wide variety of fine spectral structures are generated. Their spectral characteristics are determined by microscale physical processes involving plasma waves and non-thermal electrons that occur in the corona (e.g., Iwai et al. 2014).

Zebra patterns (ZPs) are one type of fine spectral structures. They consist of numerous, nearly parallel, drifting, narrowband stripes of enhanced emission superimposed on the background of a type IV continuum. Since type IV bursts are thought to be generated by non-thermal electrons trapped in closed magnetic structures in flaring regions, ZPs are considered to be an important source of information about the plasma parameters and processes associated with flaring loops, such as the magnetic fields and their configurations, particle acceleration, and plasma instabilities. The ZP formation mechanism has been discussed for more than 40 years, and many theoretical models have been proposed (Rosenberg 1972; Kuijpers 1975; Zheleznyakov & Zlotnik 1975; Chernov 1976; Chernov 1990; Barta & Karlicky 2006; Ledenev et al. 2006; Kuznetsov & Tsap 2007). The most sophisticated model is based on plasma wave excitation at the so-called double plasma resonance (DPR) levels (Kuijpers 1975; Zheleznyakov & Zlotnik 1975; Kuznetsov & Tsap 2007). This model assumes that the excitation of electrostatic plasma waves is enhanced at some resonance levels for which the upper-hybrid frequency $f_{UH}$ coincides with the harmonics of the electron cyclotron frequency $f_c$:

$$f_{UH} = \sqrt{f_p^2 + f_c^2} = sf_c, \qquad (1)$$

where $f_p$ is the electron plasma frequency and $s$ is the harmonic number. The emissions corresponding to different harmonic numbers originate at different heights, and the wave excitation levels are determined by the characteristic scales of the magnetic field and electron density variations. The excited plasma waves are then converted into electromagnetic waves through nonlinear processes at the fundamental plasma frequency or the second harmonic frequency. They subsequently propagate through the corona and interplanetary space and are finally observed as ZPs.

Despite the sophistication of ZP generation theories, many problems remain to be solved in this area (Chernov 2006; Zlotnik 2009). For instance, the intermediate emission polarization is an important topic of discussion. Assuming that the fundamental plasma emission is operative, the escaping radiation should be completely polarized in the ordinary mode (O-mode) in magnetoionic theory; accordingly, the degree of circular polarization (DCP) should be 100%. On the other hand, the DCP of the second harmonic emission is negligible in a weakly anisotropic plasma (Zlotnik et al. 2014). Observations have shown that the sense of rotation corresponds to the O-mode in most events but that the observed DCP is often less than 100%, though its values are rather high (Chernov et al. 1975; Chernov & Zlobec 1995). This discrepancy between the theoretical predictions and observations had led to the concept of the depolarization of fundamental plasma emission during propagation.

Several mechanisms for such depolarization have been proposed (Cohen 1960; Wentzel et al. 1986; Melrose 1989; Melrose 2006; Zlotnik et al. 2014). One mechanism is related to the mode coupling that occurs when the waves propagate through a quasi-transverse (QT) magnetic field (Cohen 1960). When the frequency of the waves is sufficiently low relative to the critical frequency, the initial mode remains and the polarization reverses, whereas at sufficiently high frequencies, the original polarization is preserved and the mode reverses. Another possible depolarization mechanism involves the contribution of weakly polarized component (Zlotnik et al. 2014). Strongly polarized emission originates from ZP stripes at the fundamental frequency $f = f_p$, while weakly polarized emission is generated by the background continuum at the second harmonic frequency from the $f_p' = f_p/2$ level. The accumulation of these components decreases the observed DCP. Furthermore, the phenomenon of large-angle scattering by lower-hybrid waves was proposed by Wentzel et al. (1986) as a depolarization mechanism of type I bursts. Melrose (1989) explored the general case of scattering by low-frequency waves and suggested that scattering by ion sound waves and whistler mode waves is a common depolarization mechanism for all metric bursts. Melrose (2006) also considered depolarization due to reflection at sharp density boundaries. Kaneda et al. (2015) investigated the circumstances under which such depolarization processes occur by analyzing the frequency dependence of the delay between the two circularly polarized components. Based on their results, they suggested that depolarization occurs very close to the emission source, where the difference between the group velocities of the two circular wave modes is relatively large.

Nevertheless, information regarding circular polarization is still lacking. Most studies concerning ZP polarization have been case studies based on the analysis of one or several events. Consequently, the general polarization characteristics of ZPs and their relations to the generation processes are not well understood, primarily due to the relatively low occurrence of ZPs, as well as the lack of instruments that can record highly resolved dynamic spectra and polarizations simultaneously.

The purpose of this research was to investigate the general polarization characteristics of ZPs in the context of their depolarization processes. To this end, we performed a statistical analysis of the polarization characteristics of ZPs using a large set of highly resolved spectral and polarization data obtained by the Assembly of Metric-band Aperture Telescope and Real-time Analysis System (AMATERAS) solar radio spectropolarimeter and developed a depolarization model that could explain the observations.

The remainder of this paper is organized as follows. In Section 2, the instrument and data set used in this study are described. Section 3 summarizes the analytical method and the results. Then, the relationship between the obtained results and depolarization processes is discussed in Section 4. Finally, we provide a summary and the main conclusions in Section 5.

## 2. DATA SET

AMATERAS is a ground-based solar radio telescope in Fukushima, Japan, that is dedicated to spectropolarimetry in the metric range (Iwai et al. 2012). It measures solar radio emission in the frequency range of 150–500 MHz with a cadence of 10 ms and a frequency resolution of 61 kHz. Simultaneous observations of the two circularly polarized components, i.e., left-handed polarization (LCP) and right-handed polarization (RCP), are possible with this telescope.

We examined the data obtained by AMATERAS between 2010 July and 2015 March. A total of 3.5 years of data are available from this period, excluding the intervals during which observations were not made. From the AMATERAS database, we selected 21 ZP events, which consisted of clearly visible narrowband stripes in 16 flares. The adjacent events that were well separated in both time and frequency and exhibited different spectral characteristics (e.g., frequency drifting or separation between the stripes) were designated as different events. These events are listed in Table 1 along with their general characteristics, including flare properties. The flare data (class, location, time of occurrence, and duration) were obtained from the *Hinode* Flare Catalog (see Watanabe et al. 2012). For each ZP event, we derived the upper- and lower-end frequencies, occurrence time, time relative to the flare peak time, duration, total number of stripes, frequency separation between adjacent stripes, and relative frequency separation. The values of the observed spectral parameters, such as the number of stripes, typical frequency separation, and overall frequency extent, that are listed in Table 1 are comparable to those observed previously (e.g., Chernov 2006). The ranges of these parameters cover almost the entire ranges reported previously. Hence, the ZP events analyzed in this study possess characteristics typical of ZPs in the metric range.

## 3. POLARIZATION CHARACTERISTICS

### 3.1. Parameters Used to Describe Polarization

The polarization of each ZP was characterized in terms of the DCP, delay between the two circularly polarized components, and magnetoionic mode of the waves. The DCP of the emission $p$ is defined as follows.

$$p = \frac{I_R - I_L}{I_R + I_L} \times 100, \tag{2}$$

where $I_R$ and $I_L$ are the radio intensities of the RCP and LCP components, respectively. The representative value for an event can be obtained by averaging over the entire time and frequency ranges of the ZP.

The O- and extraordinary (X)-modes of electromagnetic waves in magnetized plasma propagate with different group velocities; the O-mode travels faster than the X-mode. Consequently, a temporal delay between the two polarization components arises. In this study, the delay due to the difference between the group velocities was determined using the cross-correlation method (Benz & Pianezzi 1997). In this method, the cross-correlation coefficient $C(l)$ between the RCP and LCP is first calculated during a selected time interval using the following equation:

$$C(l) = \frac{\sum_{i=1}^{n-l} I_R(t_i) I_L(t_{i+l})}{\sqrt{\sum_{i=1}^{n} I_R(t_i)^2 \sum_{i=1}^{n} I_L(t_i)^2}}. \tag{3}$$

The summation in the numerator is calculated over all possible pairs for the lag $l$, and $n$ is the total number of data points in the selected time interval. Then, the calculated correlation coefficient is interpolated using cubic spline functions to increase the resolution with which the delay can be determined. The value of the lag for which the cross-correlation coefficient is maximized is then designated as the delay between the RCP and LCP components. In this study, we selected a 10 s interval during which each ZP was clearly visible. The delay was determined for each frequency channel of the ZP, and then, as in the DCP case, the delay of the event was obtained by averaging over the entire frequency range.

The magnetoionic emission mode was determined according to the temporal relation, i.e., the delay, between the two circularly polarized components. We assumed that the delay was generated during propagation because of the difference between the group velocities of the two circular wave modes. Therefore, the antecedent and following components were supposed to correspond to the O- and X-modes, respectively. In inferring the emission mode of a radio burst, the leading spot in the active region is often taken as a reference (leading spot hypothesis, Benz 1993). However, the leading spot hypothesis is not always correct and has some ambiguities. Hence, we used the delay between the two components as a reference when determining the emission mode in the present study.

*3.2. Overview of the Results*

The obtained polarization characteristics are listed in Table 2. The symbols "R" and "L" in the DCP column indicate the dominant components (e.g., R100 means 100% RCP), and those in the delay column indicate the delayed components (e.g., L50 means that the LCP component is delayed by 50 ms). Figure 1(a) presents the DCP distribution. The ZPs are weakly or intermediately circularly polarized by amounts ranging from 0% to around 70%. Among the 21 ZPs, 17 events (81.0%) have intermediate polarizations ($20\% < p < 80\%$), and 4 events (19.0%) have weak polarizations ($p \leq 20\%$). Notably, no completely polarized (100%) ZP is evident among the 21 events analyzed. The sense of the polarization remains the same throughout each flare event, though the DCP varies slightly.

A histogram of the delay between the two polarization components is provided in Figure 1(b), which shows that the delay can be as long as about 70 ms. The weaker component was delayed in 15 events (71.4%); the stronger component was delayed in three events (14.3%); and the two components were observed simultaneously in two events (9.5%). It should be noted that the delays may have been underestimated because of the frequency drift of the emission stripes. When the frequency drift is slow during the time of cross-correlation, the delay can be detected only in the frequency range of the emission stripes because delay in smooth continuum emissions cannot be detected using the employed method. Accordingly, averaging over the whole ZP frequency range would have reduced the obtained delay, though it would not have affected which component was delayed. The small delays obtained for certain events may be attributable to this effect.

The emission modes are listed in the rightmost column of Table 2. The ZP was O-mode in 17 events (81.0%) and X-mode in 3 events (14.3%). To determine the modes for the events on 2013 May 15 and 2014 May 18, for which the delay was 0 ms, we adopted the leading spot hypothesis. The magnetic polarity of the leading spot was determined using the Helioseismic and Magnetic Imager (HMI, Schou et al. 2012) onboard the *Solar Dynamics Observatory* (*SDO*). Unfortunately, the modes could not be determined definitely because the exact radio source position and magnetic field direction were unknown since spatially resolved radio observations were not obtained.

In the analyzed events, these polarization characteristics were almost constant throughout the ZP frequency range, and the dispersion within each event was relatively small compared to the variance among the events. In this paper, to focus on the differences among the events, the polarization characteristics averaged over the ZP frequency range are presented. See Kaneda et al. (2015) for a discussion of the frequency dependences of the polarization characteristics of a ZP event.

Figure 1(c) depicts the relationship between the DCP and delay. The blue, red, and green symbols represent the O-mode events, X-mode events, and event for which the DCP was 0%, respectively. The standard deviations of the parameters are indicated by the error bars. A positive correlation is evident between the DCP and delay; strongly polarized ZPs have more significant delays. Since the values are discrete and the relationship is not linear, we evaluated it by performing rank correlation. The rank correlation coefficient was 0.62, indicating a significant correlation between the two parameters.

*3.3. Examples of Typical Events*

Our results show dispersive polarization characteristics in the DCP and delay. As examples of typical events with strong circular polarization and a large delay and with weak circular polarization and a small delay, we discuss the events that occurred on 2011 June 21 and 2011 September 6 in detail.

3.3.1 Event on 2011 June 21

The event on 2011 June 21 is shown in Figure 2 as a typical example of a ZP event with strong circular polarization and a large delay. This event was analyzed in detail by Kaneda et al. (2015). A C7.7 class flare occurred in the active region of NOAA 11236 near the disk center (N14W09). It started at 01:22:00 UT, peaked at 03:25:00 UT, and ended at 04:27:00 UT. The ZP was recorded around 03:20:00 UT, just before the flare peak time. Figures 2(a) and (b) depict the dynamic ZP spectrum in terms of RCP and LCP, respectively. The ZP was enhanced intermittently during a period of about 1 minute. Figure 2(c) presents the time profiles of the RCP (blue line) and LCP (red line) at 200 MHz. The weaker LCP is clearly delayed relative to the stronger RCP. The DCP was 65% RCP, and cross-correlation analysis indicated that the delay of the LCP was 60 ms. This temporal relation suggests that the RCP was O-mode. Figure 2(d) shows the EUV image at 131 Å obtained by the Atmospheric Image Assembly (AIA) on board *SDO*. Brightening of the flaring loop structures is evident around the time that the ZP appeared. The photospheric magnetic field was obtained from the *SDO*/HMI magnetogram presented in Figure 2(e). The magnetic polarity of the leading spot was negative, so the RCP was determined to be O-mode based on the leading spot hypothesis. This finding is consistent with the inference based on the delayed component.

*3.3.2 Event on 2011 September 6*

Figure 3 depicts the event on 2011 September 6. In contrast to the event on 2011 June 21, this event exhibits weak circular polarization and a small delay. A strong type IV burst was observed around 22:30:00 UT and was associated with an X2.1 class flare that occurred in the active region of NOAA 11283 (N14W18). A bright active region is evident in the EUV image at 131 Å obtained by *SDO*/AIA, which is shown in Figure 3(d). The flare location was similar to that of the event on 2011 June 21. The flare started at 22:12:00 UT, peaked at 22:20:00 UT, and ended at 22:24:00 UT. Figures 3(a) and (b) depict the ZP observed at 22:48:00 UT in terms of RCP and LCP, respectively. It continued for about 40 s, and significant upward frequency drift is evident around 22:47:50 UT.

Figure 3(c) presents the time profiles of the two components; the blue and red lines represent the RCP and LCP, respectively. There is little difference between the time profiles of the two components. The obtained DCP was 1% RCP, and the delay of the LCP was determined to be 5 ms. Thus, the RCP was assumed to be O-mode. The magnetic polarity of the leading spot is also consistent with this statement (Figure 3(e)).

## 4. DISCUSSION

In this paper, we presented the analyses of 21 ZP events occurring during 16 flares and focused on their polarization characteristics. The polarization characteristics of the ZPs were evaluated using the DCP, delay between the two polarized components, and magnetoionic mode of the emission.

The initial polarization characteristics are determined by the emission mechanism, although they can be modulated during propagation. In the general theory of ZP polarization, the polarization characteristics are described in terms of fundamental plasma emission and depolarization during propagation as follows. The plasma frequency is generally lower than the X-mode cutoff frequency; therefore, in the source region of a metric ZP, the emission is generated only in the O-mode, and the DCP should be 100%. At this stage, the generated O-mode emission is partly converted into X-mode emission during propagation; consequently, the DCP decreases. Finally, the X-mode emission is delayed due to the difference between the O- and X-mode group velocities (Chernov & Zlobec 1995; Zlotnik et al. 2014).

Provided that an observed ZP occurs according to the above scenario, the antecedent component (corresponding to the O-mode) should be dominant, and the delayed component (corresponding to the X-mode) should be weaker. Among the 21 analyzed events, 17 events (81.0%) agree with this theory, indicating the validity of the above assumptions, whereby ZPs are generated at the fundamental frequency and depolarized during propagation.

However, the X-mode is dominant in three events (2014 February 12, 2014 November 7, and 2015 February 9), possibly due to ZP generation in X-mode, which has been discussed observationally and theoretically in the literature (Altyntsev et al. 2005; Kuznetsonv 2005). For example, Kuznetsov (2005) suggested that microwave ZPs can be generated in X-mode via nonlinear interactions between Bernstein modes. In this case, the frequency separation between adjacent stripes is expected to be equidistant with respect to the emission frequency. In fact, some events in our data set exhibited X-mode polarization and equidistant stripes (e.g., the event on 2014 February 12). Therefore, X-mode generation via Bernstein waves might be possible for metric ZPs under specific conditions wherein the magnetic field is extremely strong (several tens of Gauss).

Although the observed relation between the dominant and delayed components is almost consistent with the theory of fundamental plasma emission and subsequent depolarization, the values of DCP and delay were distributed across a wide range. In particular, the observed ZPs had DCPs ranging from very weak (near 0%) to moderately strong (about 70%) and delays ranging from 0 ms to about 70 ms. Moreover, we found a positive correlation between the DCP and delay. A possible explanation of this relationship is discussed in Section 4.1.

*4.1. Relationship between DCP and Delay*

If a source is emitting in O-mode, which is depolarized (partly converted into X-mode) during propagation, the relationship between the DCP and delay is determined by the depolarization processes. The relationship between the DCP and the depolarization is obvious: the DCP decreases as the depolarization efficiency increases. That is, the DCP is equivalent to the efficiency of conversion from O-mode to X-mode. On the other hand, the magnitude of the delay may be related to the location at which the X-mode originates because the temporal delay between the two modes arises only after depolarization occurs. The difference between the group velocities becomes infinite at the X-mode cutoff frequency and decreases as the emission frequency increases. Since the plasma frequency decreases with increasing height, the emission frequency becomes greater than the local plasma frequency as the emission propagates higher in the corona. Thus, the delay decreases as the height at which the X-mode originates increases. In other words, the delay depends on the height at which depolarization occurs. As described above, the relation between the DCP and delay is closely related to how and where depolarization occurs. We provide an explanation of the positive correlation between the DCP and delay that takes these considerations into account in Section 4.2.

*4.2. Depolarization Due to Single Reflection*

There are several depolarization models that can lead to DCP reduction. In particular, mode coupling that occurs when waves propagate through a quasi-transverse magnetic field region (i.e., a QT region) is a well-known depolarization mechanism (Cohen 1960). Zheleznyakov & Zlotnik (1964) discussed this effect and showed that the resulting DCP is strongly dependent on the radiation frequency ($\propto f^{-4}$). However, observations have indicated that the DCP is weakly frequency-dependent (Kaneda et al. 2015); this characteristic is difficult to explain using QT mode coupling theory. Furthermore, propagation through QT regions is difficult to realize for emission from coronal loops, especially from loops located near disk centers.

Zlotnik et al. (2014) proposed a mechanism based on the accumulation of emissions from different origins. According to their model, completely polarized ZP emission at the fundamental frequency $f = f_p$ and weakly polarized continuum emission at the second harmonic frequency from the $f_p' = f_p/2$ layer are observed in the same frequency band. As a result, the total emission can be partly polarized. This mechanism may not be unnatural in coronal conditions, but it cannot explain our observations. If the depolarization were attributable to the contribution from weakly polarized emission, ZP stripes would be observable only in one polarized component. However, we observed ZP stripes in both polarized components. Thus, some other mechanisms should be associated with the observed depolarization.

Another depolarization mechanism involves scattering by low-frequency waves such as lower-hybrid waves

(Wentzel et al. 1986), ion sound waves, or whistler mode waves (Melrose 1989). The occurrence of such scattering requires the existence of low-frequency waves that interact with the initial radiation.

As another possible depolarization mechanism that can explain the observed polarization characteristics, we consider a model based on reflection at sharp plasma density boundaries. A simplified theory of depolarization due to reflection was proposed by Melrose (2006). According to this theory, when a wave in one magnetoionic mode is reflected at a sharp boundary, waves in both modes are generated. As a result, the DCP is reduced. Such reflection requires a boundary with a thickness less than about the wavelength of the radio waves. Melrose (2006) derived a depolarization coefficient using simplified magnetoionic theory. In this model, the magnetic field is ignored in the refractive indices for the magnetoionic modes but is taken into account in the circular polarization. Under this assumption, the RCP and LCP reflection coefficients are treated as equivalent, i.e., $r_{RR} = r_{LL}$ and $r_{RL} = r_{LR}$, where the first and second subscripts represent the polarizations of the incident and reflected modes, respectively. Assuming incident RCP, the depolarization coefficient can be written as

$$p_R = \frac{r_{RR} - r_{RL}}{r_{RR} + r_{RL}}, \tag{4}$$

and the detailed depolarization coefficient expressions are

$$p_R = \frac{\cos 2\theta + \cos 2\theta'}{1 + \cos 2\theta \cos 2\theta'}, \quad \cos 2\theta' = \frac{X' - X + (1 - X)\cos 2\theta}{1 - X'}, \tag{5}$$

for $X < X' < 1 - (1 - X)\cos^2 \theta$, and

$$p_R = \frac{x - (1 + 3x)\sin^2 \theta + 2x \sin^4 \theta}{x + (1 - x)\sin^2 \theta}, \quad x = \frac{1 - X}{X' - 1}, \tag{6}$$

for $1 - (1 - X)\cos^2 \theta < X'$. In this context, $X = \omega_p^2/\omega^2$ specifies the ratio of the plasma frequency to the wave frequency on the lower density side, $X' = \omega_p'^2/\omega^2 = \xi X$ is the corresponding ratio on the higher density side, $\theta$ is the angle between the magnetic field lines and incident wave vector, and $\theta'$ is the corresponding angle of the reflected wave. For a detailed derivation of $p_R$, see Melrose (2006).

The calculated depolarization coefficient as a function of $\theta$ and $\omega/\omega_p$ is shown in Figure 4 (left panel), where $\theta$ is the angle between the incident ray and magnetic field and $\omega/\omega_p$ is the ratio of the emission frequency to the local plasma frequency. Here, the DCP becomes 0% at $p_R = 0$, there is no change in the DCP at $p_R = 1$, and complete polarization reversal occurs at $p_R = -1$. The depolarization coefficient is close to 1 when $\theta \lesssim 5°$ and $\omega/\omega_p \gtrsim 2$, and it is close to −1 when $\theta \gtrsim 30°$ and $\omega/\omega_p \lesssim 1.2$.

The right panel of Figure 4 depicts the frequency dependence of the depolarization coefficient at $\theta = 5°$. Since $\omega/\omega_p$ increases with increasing height, the frequency dependence of the depolarization coefficient implies that the DCP reduction becomes less significant as the reflection height increases. Consequently, the DCP and delay should be negatively correlated if they result from a single depolarizing reflection. However, the observational results conflict with this prediction, since the DCP and delay were in fact positively correlated. Thus, single reflections with different heights may not be the reason for the observed relationship between the DCP and delay.

*4.3. Depolarization Due to Multiple Reflections*

Alternatively, the dispersive polarization characteristics may be related to the multiplicity of depolarization processes, which is highly likely. In the case of depolarization due to reflection at sharp boundaries, the DCP and delay depend on the number of reflections. In the model considered, it is possible for the polarization to change from O-mode to X-mode and vice versa due to reflection. Thus, if the incident wave is a mixture of the O- and X-modes, four kinds of reflected waves are produced by one reflection: the O- and X-modes reflected from the incident O-mode (O-O and O-X), and those reflected from the incident X-mode (X-O and X-X). In the observations, the accumulation of these four components was received as the two polarized components. Accounting for this accumulation, the time profiles of the observed O- and X-mode components after *i*-th reflections can be expressed as

$$O_i(t) = \frac{1+p_{Ri}}{2} O_{i-1}(t) + \frac{1-p_{Ri}}{2} X_{i-1}(t+\Delta t_i) \tag{7}$$

and

$$X_i(t) = \frac{1+p_{Ri}}{2} X_{i-1}(t) + \frac{1-p_{Ri}}{2} O_{i-1}(t-\Delta t_i), \tag{8}$$

where $p_{Ri}$ is the depolarization coefficient in equation (4) at the *i*-th reflection and $\Delta t_i$ is the delay generated during propagation from the height at which the *i*-th reflection occurs. Due to the contribution from the oppositely polarized component generated upon the $(i-1)$-th reflection, the O- and X-modes are mixed. As the number of reflections increase, the mixing of the two components proceeds, and, consequently, the DCP and delay decrease.

Figure 5 illustrates this model schematically. Typical ZP time profiles are presented for cases with zero, one, and two reflections, from top to bottom. The blue and red lines represent the O- and X-modes, respectively. Here, we assumed that the originally generated ZP was completely polarized in the O-mode, corresponding to the zero-reflection case in the upper panel. Upon the first reflection, a fraction of the incident O-mode is transformed into the X-mode according to $p_{R1}$, and the generated X-mode is delayed by $\Delta t_1$ (middle panel). After the second reflection, there are four reflected components, as shown in the lower panel: O-O (blue dotted), O-X (red dotted), X-O (blue dashed), and X-X (red dashed). The observed O-mode time profile (blue solid) is represented as the sum of the O-O and X-O components, and the X-mode profile (red solid) is represented as the sum of the O-X and X-X components. The mixing of the two modes makes the O-mode to X-mode intensity ratio close to 1; thus, the DCP decreases. Furthermore, since the four reflected components have different delays, the overall delay between the O- and X-modes is lower than that in the one-reflection case. Consequently, the DCP and delay decrease as the number of reflections increases.

*4.4. Comparison between Model and Observations*

To examine the relationship between the polarization characteristics and number of reflections, we performed calculations using a simplified reflection model. We assumed an artificial time profile as the originally generated ZP emission in the source and calculated variations in polarization characteristics due to depolarization. Then, the DCP and delay were derived from the resulting time profiles of the O- and X-modes. We considered depolarization due to the reflections and the difference between the group velocities of the two magnetoionic modes. Furthermore, the calculated polarization characteristics were compared with the observational results.

To calculate the depolarization coefficient, we used the model of Melrose (2006). In this model, (1) the magnetic field is ignored in the refractive indices but is taken into account in the circular polarizations; (2) the sharp boundary layer is assumed to be parallel to the magnetic field lines and to have a thickness of less than about one wavelength; and (3) the incident waves, magnetic field, and normal to the boundary are assumed to be coplanar. The expressions for the depolarization coefficient are given in equations (5) and (6).

The expected delay $\Delta t$ between the two modes at a frequency $\omega$ can be described as follows.

$$\Delta t = \int_{s_0}^{\infty} \left( \frac{1}{v_X} - \frac{1}{v_O} \right) ds, \tag{9}$$

where $v_O$ and $v_X$ are the group velocities of the O- and X-mode waves, respectively, and $s$ is the coordinate along the propagation path from the reflection height $s_0$ to the observer (Yasnov & Karlicky 2003). In this study, the integral in equation (9) was performed along the radial direction. Accounting for the changes of the plasma parameters along the radial direction, we used density and magnetic field models of the corona, i.e., the 10-fold Baumbach–Allen model for the electron density (Allen 1947) and the Dulk & Mclean model for the magnetic field (Dulk & Mclean 1978).

Additionally, the multiplicity of reflections is included in the proposed model. The reflection height increases with each reflection. For simplicity, the interval of each reflection is assumed to be constant: reflection occurs every time $\omega/\omega_p$ increases by 0.05. The first reflection is placed at $\omega/\omega_p = 1.1$. In the general coronal model, $\omega$ is higher than the X-mode cutoff frequency $\omega_X$ for these parameters, and, hence, both modes can propagate there. The density ratio $\xi = 10$ and angle $\theta = 5°$ are also assumed to be constant throughout all reflections. We used a sine curve with a period of 1.2 s as an initial time profile representing typical ZP emission. The expressions in equations (7) and (8) were used to calculate the time profiles of the two polarized components after *i* reflections. The calculated time profiles with 0, 1, and 12 reflections are presented in Figure 6, from top to bottom. The blue and red lines correspond to the O- and X-modes, respectively. If there are no reflections during propagation, depolarization does not occur and the originally generated O-mode is preserved. Accordingly, the observed DCP should be 100% and the delay cannot be detected, as shown in the upper panel of Figure 6. In the case of one reflection (middle panel), the DCP and delay are determined by the reflection height. One reflection at $\omega/\omega_p = 1.1$ leads to a DCP of 50% and a delay of 60 ms. Reflections closer to the source cause significant depolarization and delays, as expected. When the number of reflections increases to 12, the DCP and delay decrease to 6% and 6 ms, respectively (lower panel).

These results suggest that the positive correlation between the DCP and delay can be explained by the difference between the numbers of reflections, though the possibility of depolarization due to scattering by low-frequency waves cannot be excluded. Among the 21 events analyzed in this study, no event had a DCP close to 100%. This finding implies that each observed ZP involved at least one reflection during propagation through the corona. However, past observations showed that the DCPs of ZPs can be as high as 100% (Chernov 2006; Tan et al. 2014). The reason for this discrepancy is difficult to specify, though the small sample size used in this research may be the reason. As shown in Figure 4, the depolarization coefficient is heavily dependent on $\theta$ and $\omega/\omega_p$. In fact, these parameters can vary widely under coronal conditions. Nevertheless, the assumptions made about the angle and frequency in the calculations seem plausible for developing a general understanding of multiple reflections for the following reasons.

First, the width of the angular spectrum is confined to small angles nearly parallel to the density gradient due to the effects of refraction. The maximum escape angle $\varphi_{max}$ of the radiation is given by the following formula (Zheleznyakov 1996):

$$\varphi_{max} = \text{arcsec}\frac{\omega}{\omega_p}. \tag{10}$$

Assuming the DPR model, then

$$\frac{\omega}{\omega_p} = \frac{s}{\sqrt{s^2 - 1}}. \tag{11}$$

Using equations (10) and (11), the maximum escape angles for typical ZP stripe harmonic numbers can be obtained: $\varphi_{max} = 5.7°$ for $s = 10$ and $\varphi_{max} = 2.9°$ for $s = 20$. These values are almost consistent with those calculated in this study. Moreover, the tendency for the DCP and delay to decrease as the number of reflections increases does not change with the angle. Figure 7 shows the angular dependence of the polarization change due to multiple reflections. The upper and lower panels depict the DCP and delay, respectively. In each panel, cases with $\theta = 3°$, $4°$, $5°$, $6°$, and $7°$ are represented by purple, blue, green, orange, and red lines, respectively. Although the two parameters decrease in different ways for different angles, they monotonically decrease as the number of reflections increases.

Meanwhile, $\omega/\omega_p$ is determined by the relation between the heights at which the radiation originates and reflection occurs. In DPR conditions, the sources are distributed over a wide range of heights, and reflection points cannot be specified. Accordingly, $\omega/\omega_p$ can also be dispersive, but it still increases as the waves propagate outward along the density gradient, because $\omega_p$ decreases with height. Although the assumption that the waves are reflected in a constant frequency interval may be too much of a simplification, the essential feature that $\omega/\omega_p$ increases with the number of reflections can still be obtained. In any case, more detailed modeling will be necessary for further discussion of the relation between depolarization and the observed polarization characteristics, such as the constancy of the DCP and delay with respect to the time and frequency.

*4.5. Picture of the Source Region*

For depolarizing reflection to occur, a sharp density gradient whose thickness is less than the wavelength, which is on the order of meters in this case, is necessary. The formation of such a boundary is closely related to the microscopic structure of the corona. Due to the recent development of imaging instruments with high resolutions in the EUV wavelength range, such as *SDO*/AIA and *HINODE*/EIS, the fine structure of the corona has been revealed (e.g., Del Zanna 2008; Van Doorsseleaere et al. 2008; Warren et al. 2008; Brooks et al. 2012; Peter et al. 2013; Scullion et al. 2014).

The EUV observations indicate that coronal loops consist of bundles of thin strands. The minimum size of such strands is estimated to be less than 100 km (Peter et al. 2013; Scullion et al. 2014), and each strand has different physical properties, including density (Van Doorsseleaere et al. 2008; Brooks et al. 2012), temperature (Warren et al. 2008), and flow speed (Del Zanna 2008). At the boundaries between strands with different physical properties, tangential discontinuities can easily be formed. The thickness of such a discontinuity is determined by the ion gyroradius. Under typical conditions in the coronal active region (a magnetic field $B = 10$ G, and a temperature $T = 10^6$ K), the ion gyroradius is about 1 m, which is sharp enough to cause reflection and, accordingly, depolarization to occur. In addition, the tangential discontinuity is stable against spatial diffusion in the direction perpendicular to the magnetic field. For the spatial scale of 10 m, the diffusion time is on the order of 100 s (at $T = 10^6$ K), which is longer than the typical durations of ZPs (several tens of seconds). Therefore, the existence of few-meter-wide boundaries that are stable for longer than ZP lifetimes seems to be consistent with the microscale structure of the corona inferred from high-resolution imaging observations.

Two different methods of reflection in the source region were considered by Melrose (2006) for fundamental emission in type I, II, and III bursts: reflection off bundles of overdense fibers for type I bursts and reflection off the walls of ducts for type II and III bursts. The polarization of a type I burst is known to depend on the location of the source region (Zlobec 1975; Wentzel et al. 1986). The DCP is high when the source is located near the central meridian, and it becomes low as the source approaches the limb. This longitudinal dependence is interpreted as resulting from the occurrence of large-angle reflections in overdense fibrous structures (Bougeret & Steinberg 1977; Wentzel et al. 1986). In spite of the similarity of the source geometries (closed-loop structures) of type I and IV bursts, such longitudinal dependence was not observed for the ZPs analyzed in this study. For example, the events on 2011 June 21 and 2011 September 6 occurred at almost the same longitude and had similar active region configurations, but their polarization characteristics were quite different.

Furthermore, it is difficult to explain the multiplicity of depolarizing reflections in the context of large-angle reflections, because the reflection coefficient becomes very small as $\omega/\omega_p$ increases and most of the energy is transmitted. As a result, effective depolarization does not occur in this case. Thus, reflection off the bundles of overdense fibrous structures may not be plausible in ZP source regions. Alternatively, ducting along low-density flux tubes with overdense walls can reasonably explain the multiplicity of reflections (Duncan 1979). For ducting to occur, low-density open field lines must exist around the source, though the sources of ZPs are considered to be hot flaring loops. Hence, the closed-loop structure, which is the emission source, and low-density flux tubes with overdense walls should be adjacent to each other for ducting to occur. Such conditions can be realized near the footpoints of complicated magnetic structures.

The picture of ZP generation and propagation suggested by our results can be described as follows. The emission itself is generated at the fundamental plasma frequency in a flaring loop structure filled with non-thermal electrons. Then, the emission propagates through a low-density background and is reflected by overdense loops near the source region. A schematic diagram of the expected source region is provided in Figure 8. It should be noted that the existence of overdense structures with sharp boundaries near the source can also cause the conversion of electrostatic waves into electromagnetic waves.

We suggested that the differences between the polarization characteristics can be explained by the differences between the numbers of reflections, assuming fundamental plasma emission. Nevertheless, what determines the number of reflections is not clear. As one possibility, we suggest the fine-scale structures of coronal loops. As mentioned above, coronal loops are considered to be collections of thin strands. The different compositions of such strands within a loop could result in different wave ducting paths and change the number of reflections. Moreover, in the low corona where the fine structures exist, the magnetic field is relatively strong. In such a strong magnetic field, the generation of ZPs in the X-mode becomes possible (Kuznetsov 2005), which may be related to the observed X-mode-dominant events. Unfortunately, we have no spatial information regarding the radio emission sources, so it is difficult to identify the differences between the fine structures of the loops in the sources.

## 5. CONCLUSIONS

We investigated the polarization characteristics of 21 selected ZP events observed by AMATERAS. Although the number of events was limited, the set of events enabled us to understand the general polarization characteristics of ZPs. The results can be summarized as follows. (1) The ZPs were weakly or moderately circularly polarized, and the DCP was distributed across a wide range from near 0% to about 70%. (2) A delay of up to 70 ms was detected between the two circularly polarized components. For most events, the weaker polarization component was delayed. (3) The emission mode of the waves was determined based on the delayed component and was O-mode in about 81% of the events.

Most of the events could be interpreted in the framework of fundamental plasma emission and depolarization during propagation. More specifically, the emission was originally generated and polarized in the O-mode and was then depolarized and partly converted into the X-mode. The difference between the group velocities of the two modes caused the delay of the X-mode (i.e., the weaker component).

We found a positive correlation between the DCP and delay. As a possible depolarization mechanism, a model based on multiple reflections at sharp boundaries was considered. We calculated the expected polarization characteristics after multiple reflections and compared them with the observational results. The results showed that depolarization due to multiple reflections can produce the observed polarization characteristics: as the number of reflections increases, the DCP and delay decrease. This finding implies that the polarization characteristics of ZPs can be determined by their numbers of reflections. However, the obtained agreement between the model and observations is still qualitative. For a quantitative discussion, including the angular and frequency dependences of depolarization, it is necessary to use more accurate models of the plasma density, magnetic field, and source geometry. Such models can be developed by performing high-resolution radio imaging observations using multi-frequency radioheliographs such as the Chinese Spectral Radio Heliograph (Yan et al. 2009).

What determines the number of reflections cannot be specified without spatial information about the radio source. One possible cause is related to the fine-scale structures of coronal loops with different physical properties. To identify the differences between the fine structures of loops, coordinate observations at radio and other wavelengths with high spatial and temporal resolutions are essential.

*Acknowledgements*

AMATERAS is a Japanese radio telescope developed and operated by Tohoku University. The *SDO* data were supplied courtesy of NASA/*SDO* and the AIA and HMI science teams. This work was partly carried out by using *Hinode* Flare Catalog (http://st4a.stelab.nagoya-u.ac.jp/hinode_flare/), which is maintained by ISAS/JAXA and Solar-Terrestrial Environment Laboratory, Nagoya University. This work was conducted by a joint research program of the Institute for Space-Earth Environmental Research, Nagoya University.

**FIGURES and TABLES**

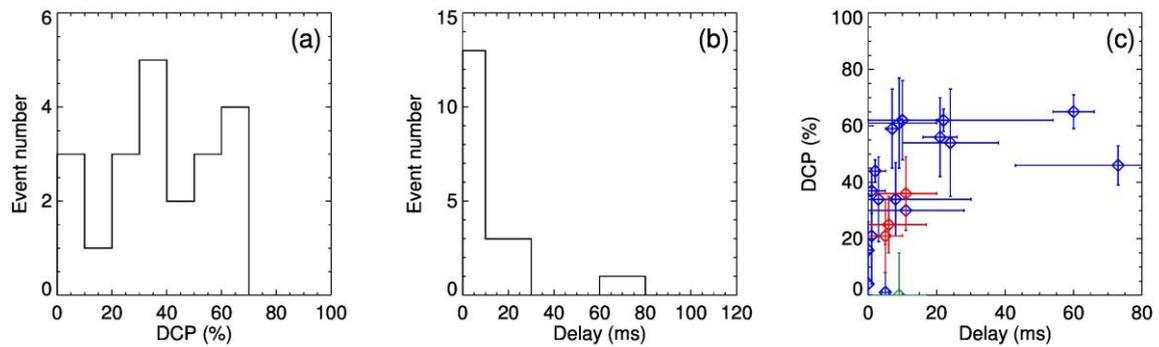

**Figure 1.** Histograms of (a) the DCP and (b) the delay between the two polarization components, not accounting for the polarization sense. (c) Scatter plot of DCP vs. delay. The blue, red, and green symbols represent the O-mode, X-mode, and not circularly polarized events, respectively. The rank correlation coefficient was determined to be 0.62.

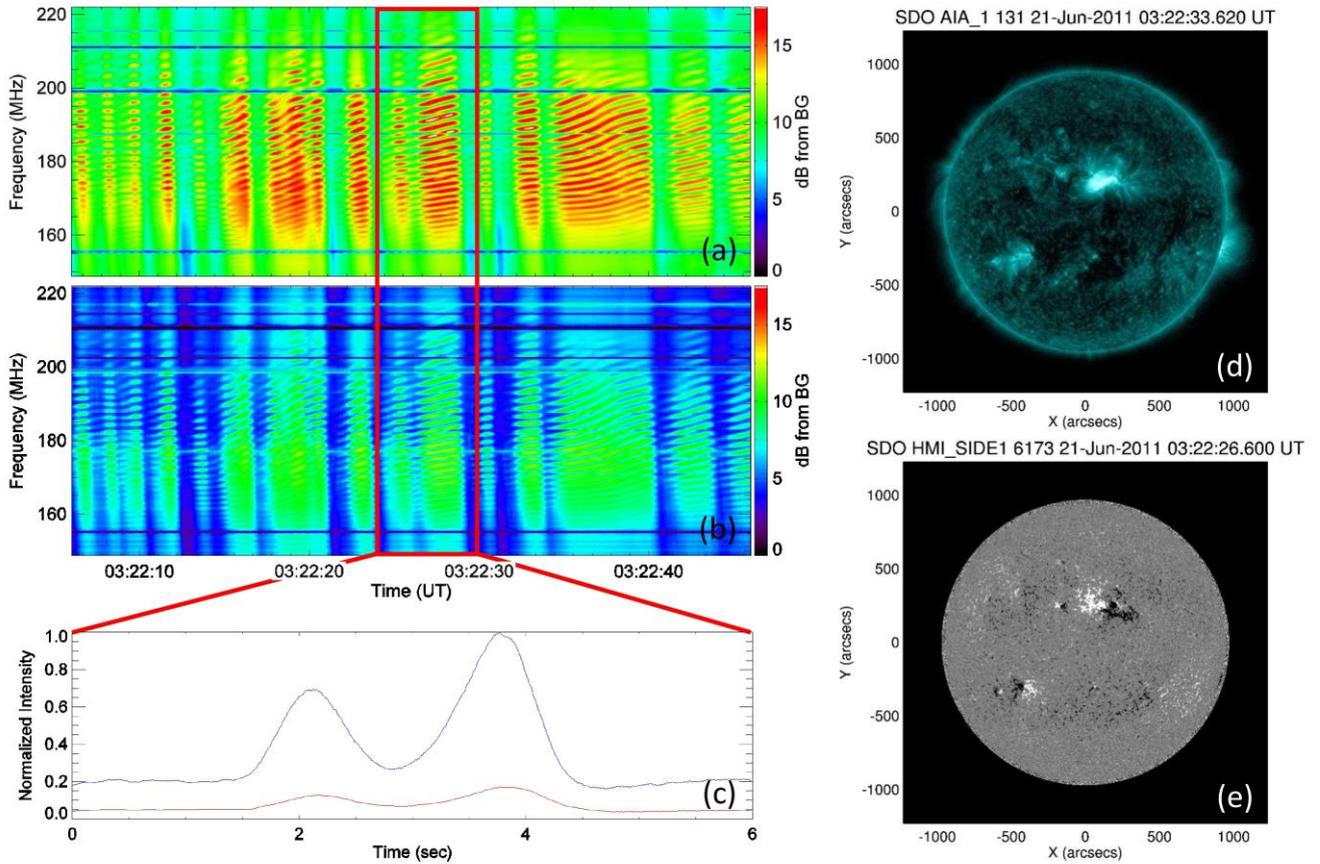

**Figure 2.** Event on 2011 June 21. Dynamic spectra of ZPs in (a) RCP and (b) LCP from 03:22:06 UT until 03:22:46 UT. (c) Time profiles of the RCP (blue line) and LCP (red line) at 200 MHz in the time range indicated by the red rectangle (i.e., 03:22:24–03:22:30 UT). (d) Full disk image (131 Å) taken at 03:22:33 UT by *SDO*/AIA. (e) Full disk magnetogram obtained by *SDO*/HMI at 03:22:26 UT.

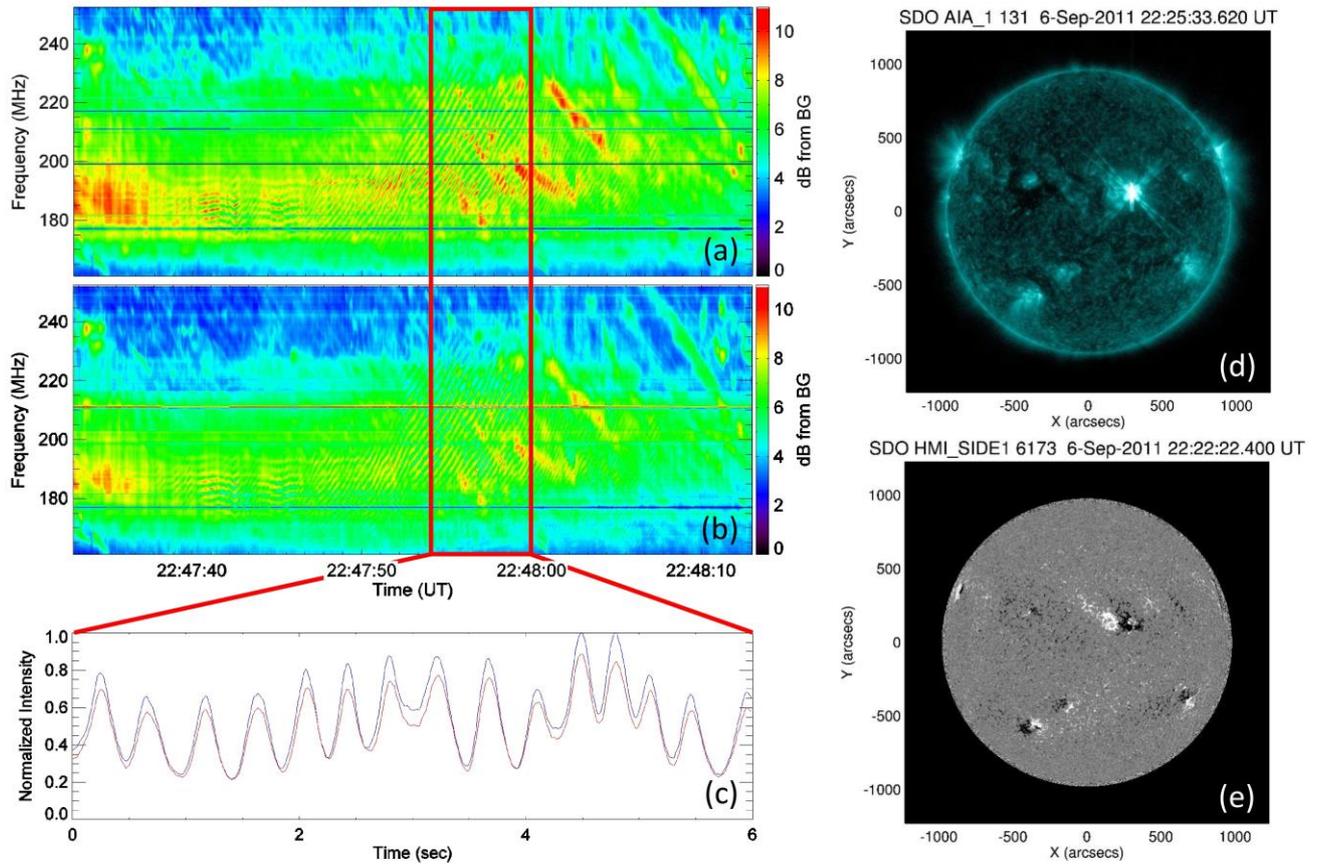

**Figure 3.** Event on 2011 September 6. Dynamic spectra of ZPs in (a) RCP and (b) LCP from 22:47:33 UT until 22:48:13 UT. (c) Time profiles of RCP (blue line) and LCP (red line) at 210 MHz in the time range indicated by the red rectangle (i.e., 22:47:54–22:48:00 UT). (d) Full disk image (131 Å) taken at 22:25:33 UT by *SDO*/AIA. (e) Full disk magnetogram obtained by *SDO*/HMI at 22:22:22 UT.

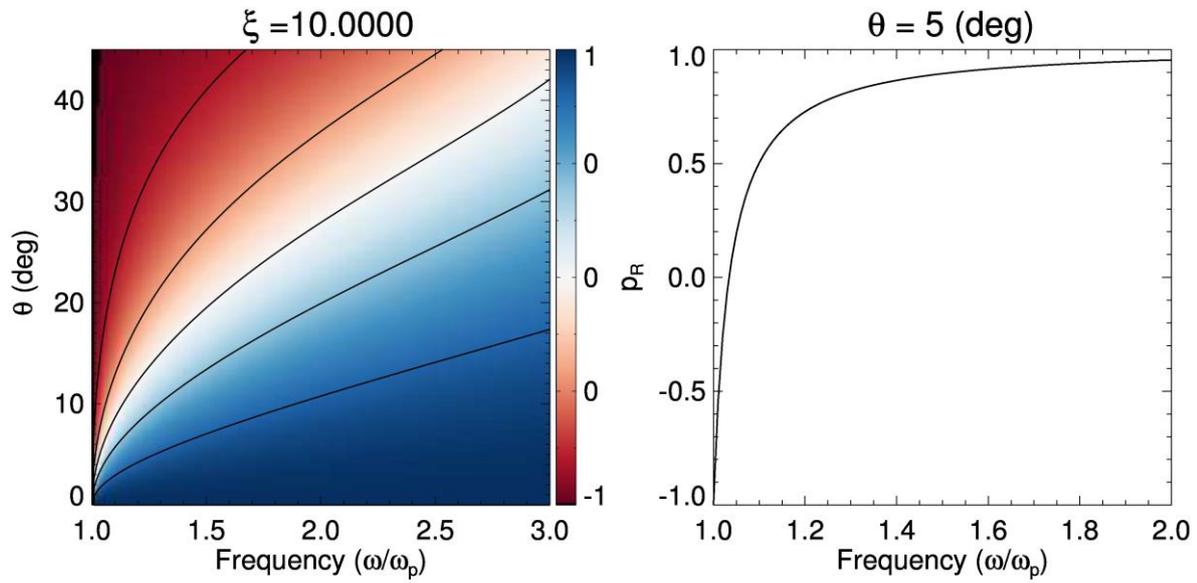

**Figure 4.** Left: depolarization coefficient calculated using equations (5) and (6) for $\xi = \omega_p'^2/\omega^2 = 10$, after Melrose (2006). The contours of 0.8, 0.4, 0, -0.4, and -0.8 from top to bottom are shown. Right: frequency dependence of depolarization coefficient when $\theta = 5°$.

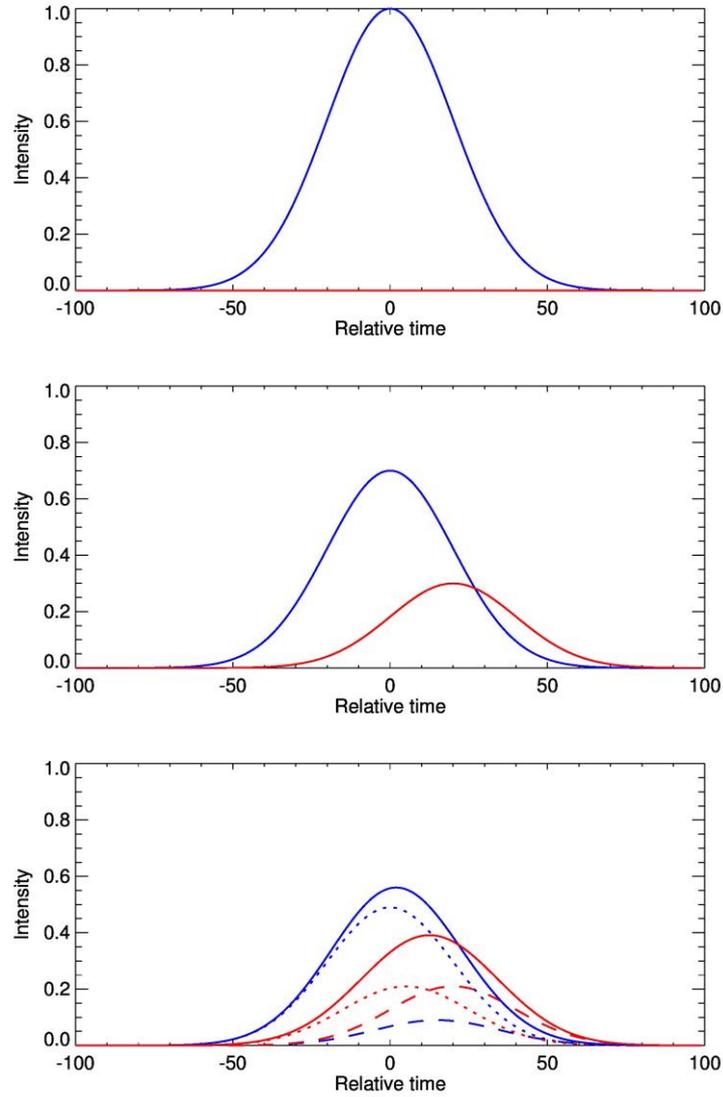

**Figure 5.** Schematic diagrams showing the effects of multiple reflections. The cases with zero, one, and two reflections are depicted in the upper, middle, and lower panels, respectively. The blue and red lines correspond to the O- and X-modes, respectively. The dotted and dashed lines in the lower panel represent the reflected components from the incident O- and X-modes, respectively.

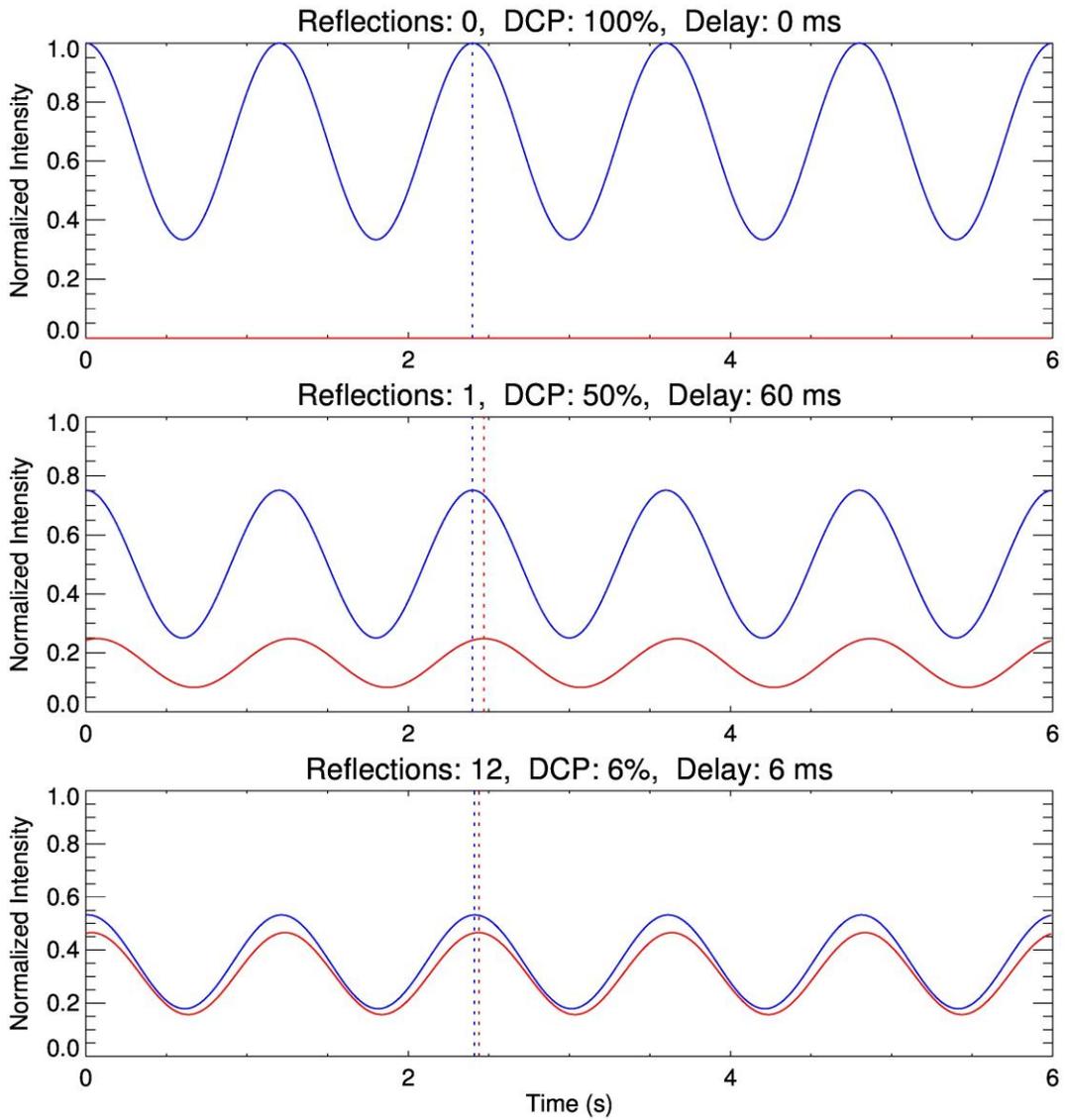

**Figure 6.** Calculated time profiles of both modes with 0, 1, and 12 reflections, from top to bottom. The blue and red lines represent the O- and X-modes, respectively. See the text for the parameters used in the calculations.

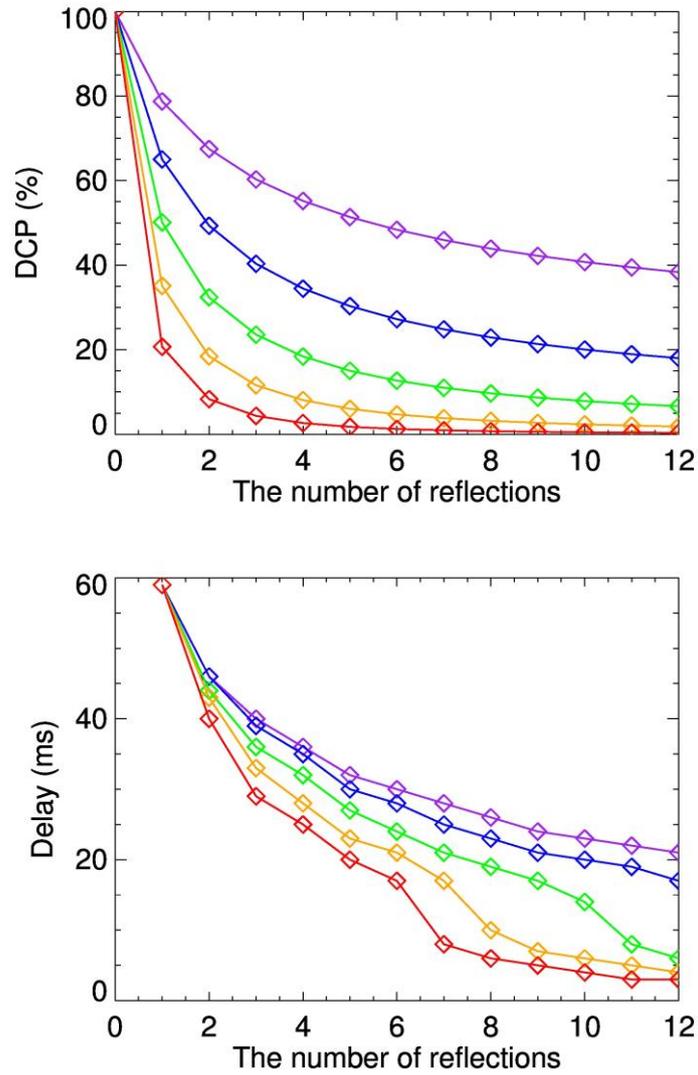

**Figure 7.** Dependences of the DCP (upper panel) and delay (lower panel) on the number of reflections for waves with different incident angles. The purple, blue, green, orange, and red lines corresponds to $\theta = 3°$, $4°$, $5°$, $6°$, and $7°$, respectively. The other values used in the calculations are the same as those used to obtain Figure 6.

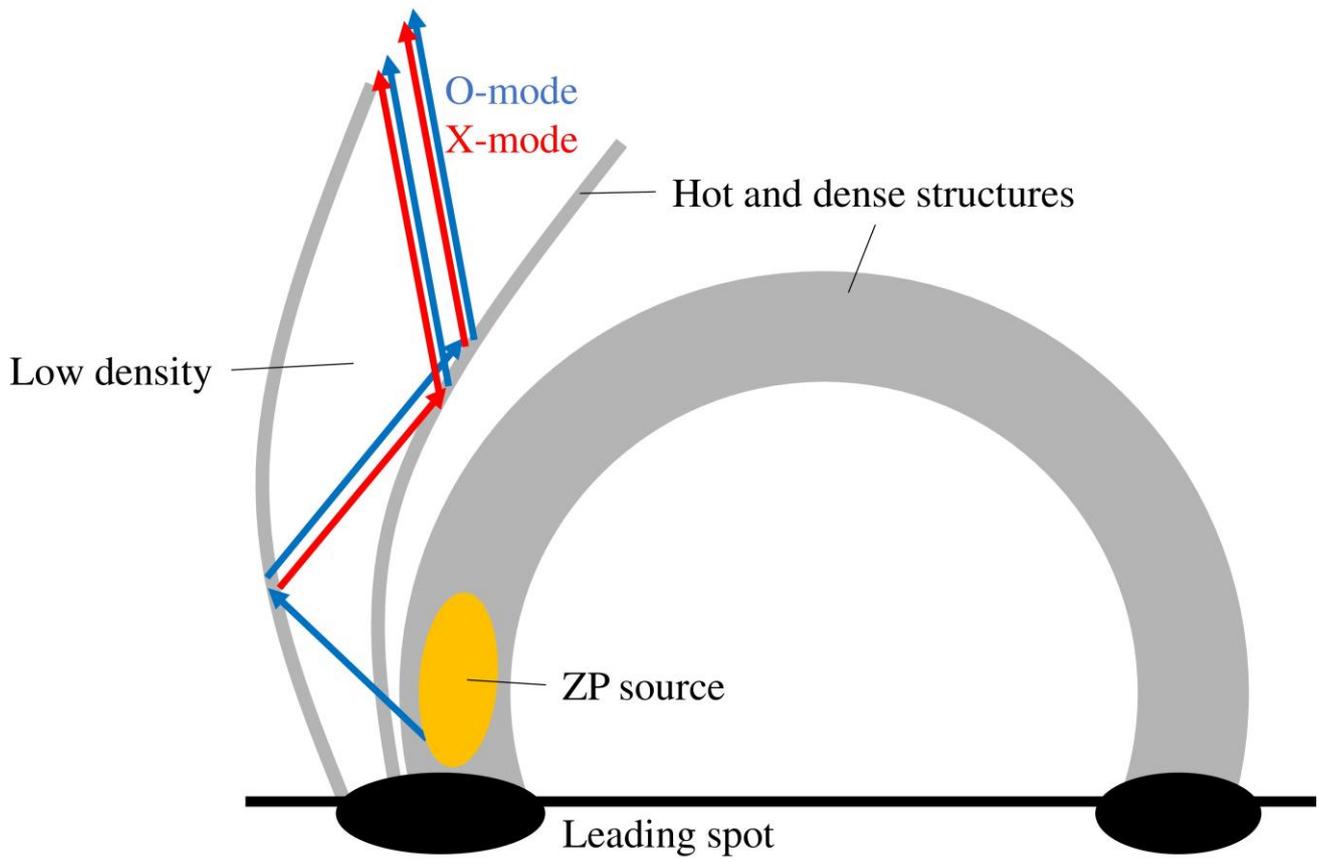

**Figure 8.** Schematic diagram of ZP generation and propagation. The blue and red arrows represent the O- and X-modes, respectively.

**Table 1.** ZP events and their general characteristics.

| Date | Class | Source Location | $t_{st}$ (UT) | $t_{fp}$ (UT) | $t_{ed}$ (UT) | $D_{fr}$ (minutes) | $t_{ZP}$ (UT) | $t_r$ (minutes) | $f_l$ (MHz) | $f_h$ (MHz) | N | $D_{ZP}$ (s) | $\Delta f$ (MHz) | $\Delta f/f$ |
|---|---|---|---|---|---|---|---|---|---|---|---|---|---|---|
| 2011 Jun 2 | C1.4 | S26E12 | 06:30 | 06:31 | 06:36 | 6 | 06:34 | 3 | 220 | 260 | 15 | 4 | 2.6 | 0.011 |
| 2011 Jun 21 | C7.7 | N14W09 | 01:22 | 03:25 | 04:27 | 185 | 03:22 | -3 | 155 | 215 | 31 | 55 | 2.0 | 0.011 |
| 2011 Aug 2 | M1.4 | N17W12 | 05:19 | 06:19 | 06:48 | 99 | 06:24 | 5 | 300 | 390 | 15 | 120 | 5.4 | 0.016 |
|  |  |  |  |  |  |  | 06:25 | 6 | 155 | 205 | 10 | 25 | 5.7 | 0.032 |
|  |  |  |  |  |  |  | 06:32 | 13 | 245 | 340 | 30 | 30 | 3.3 | 0.011 |
| 2011 Sep 6 | X2.1 | N14W18 | 22:12 | 22:20 | 22:24 | 12 | 22:48 | 28 | 175 | 230 | 12 | 40 | 3.7 | 0.018 |
| 2013 May 15 | X1.2 | N12E64 | 01:24 | 01:40 | 02:30 | 66 | 01:36 | -4 | 180 | 240 | 6 | 50 | 2.2 | 0.011 |
| 2013 Nov 11 | C7.8 | S13W24 | 00:43 | 00:48 | 00:51 | 8 | 00:45 | -3 | 230 | 270 | 5 | 90 | 6.0 | 0.024 |
| 2014 Feb 11 | C2.5 | S11E17 | 00:10 | 01:03 | 01:18 | 68 | 01:10 | 7 | 220 | 320 | 15 | 450 | 3.6 | 0.014 |
|  | M1.7 | S12E17 | 03:14 | 03:31 | 04:00 | 46 | 04:05 | 34 | 430 | 475 | 6 | 15 | 4.7 | 0.01 |
| 2014 Feb 12 | M3.7 | S12W02 | 03:52 | 04:25 | 04:38 | 46 | 05:14 | 49 | 225 | 310 | 21 | 40 | 2.9 | 0.012 |
| 2014 Apr 3 | C5.5 | N10E50 | 04:04 | 04:14 | 04:26 | 22 | 04:21 | 7 | 165 | 185 | 8 | 2 | 2.6 | 0.015 |
| 2014 May 18 | C1.2 | N12E05 | 02:33 | 02:39 | 02:45 | 12 | 02:37 | -2 | 140 | 200 | 17 | 40 | 4.0 | 0.022 |
| 2014 Jun 12 | M3.1 | S20W65 | 21:34 | 22:16 | 22:52 | 78 | 22:09 | -7 | 230 | 320 | 13 | 85 | 5.9 | 0.021 |
|  |  |  |  |  |  |  | 22:17 | 1 | 175 | 300 | 18 | 25 | 3.5 | 0.014 |
| 2014 Nov 6 | M5.4 | N17E58 | 03:29 | 03:50 | 05:12 | 103 | 04:01 | 11 | 170 | 190 | 4 | 35 | 2.5 | 0.014 |
| 2014 Nov 7 | M2.7 | N17E50 | 02:01 | 02:44 | 05:51 | 230 | 03:20 | 36 | 235 | 410 | 27 | 110 | 3.5 | 0.01 |
| 2014 Nov 7 | M2.0 | N17E50 | 04:12 | 04:25 | 04:38 | 26 | 04:12 | -13 | 150 | 185 | 14 | 250 | 1.5 | 0.009 |
|  |  |  |  |  |  | 26 | 04:21 | -4 | 150 | 210 | 18 | 280 | 2.5 | 0.014 |
| 2015 Feb 9 | M2.4 | N12E61 | 22:59 | 23:35 | 00:14 | 75 | 23:27 | -8 | 220 | 250 | 11 | 35 | 2.7 | 0.011 |
|  |  |  |  |  |  |  | 23:41 | 6 | 220 | 260 | 6 | 50 | 5.4 | 0.023 |

**Note.** $t_{st}$: flare start time; $t_{fp}$: flare peak time; $t_{ed}$: flare end time; $D_{fr}$: flare duration; $t_{ZP}$: central time of ZP; $t_r$: relative time of ZP with respect to the flare peak time; $f_l$ and $f_h$: lower- and higher-end frequencies of ZP, respectively; $N$: number of stripes; $D_{ZP}$: ZP duration; $\Delta f$: frequency separation between adjacent stripes; $\Delta f/f$: relative frequency separation with respect to the central ZP frequency. Although we obtained data from 100 MHz to 500 MHz, the sensitivity was low at frequencies less than 150 MHz, and the data exhibited strong interference due to artificial signals at frequencies greater than 480 MHz.

**Table 2.** Polarization characteristics of ZPs.

| Date | DCP (%) | Delay (ms) | Mode |
|---|---|---|---|
| 2011 Jun 2 | L34 | R3 | O |
| 2011 Jun 21 | R65 | L60 | O |
| 2011 Aug 2 | L37 | R1 | O |
|  | L21 | R1 | O |
|  | L62 | R22 | O |
| 2011 Sep 6 | R1 | L5 | O |
| 2013 May 15 | L4 | 0 | O |
| 2013 Nov 11 | L61 | R9 | O |
| 2014 Feb 11 | R59 | L7 | O |
|  | 0 | L9 | - |
| 2014 Feb 12 | L21 | L5 | X |
| 2014 Apr 3 | L34 | R8 | O |
| 2014 May 18 | R16 | 0 | O |
| 2014 Jun 12 | L44 | R2 | O |
|  | L56 | R21 | O |
| 2014 Nov 6 | L30 | R11 | O |
| 2014 Nov 7 | R62 | L10 | O |
| 2014 Nov 7 | L25 | L6 | X |
|  | L54 | R24 | O |
| 2015 Feb 9 | R46 | L73 | O |
|  | R36 | R11 | X |